\documentstyle{article}

\renewcommand{\thefootnote}{\fnsymbol{footnote}}
\begin{document}
\thispagestyle{empty}
\hfill{hep-th/9904192}
\vspace{0.5cm}
\begin{center}
{\Large {\bf Massive $4d$ particle with torsion and conformal mechanics}}\\
\vspace{0.5cm}
{\large A.Nersessian}\footnote{e-mail:nerses@thsun1.jinr.ru}\\

 {\it Laboratory of Theoretical Physics, JINR,
 Dubna, 141980, Russia}\footnote{Address for correspondence}\\

and\\

 {\it Department of Theoretical Physics, Yerevan State University, \\
A. Manoukian St., 5, 375012, Yerevan, Armenia}\\
\end{center}

\begin{abstract}
The consequences of coupling of the torsion (highest curvature)
term to the Lagrangian of a
massive spinless particle in four-dimensional space time are
 studied.
It is shown that the modified system remains spinless and possesses
 extended gauge invariance.
Though the torsion term does not generate spin,
it provides the system with a nontrivial mass spectrum,
described by one-dimensional conformal mechanics.
Under an appropriate choice of characteristic constants the system has
  solutions with a discrete mass spectrum.
\end{abstract}
\newpage
\renewcommand{\thefootnote}{\arabic{footnote}}
\setcounter{footnote}0
\section{Introduction}

The search of Lagrangian models, describing spinning particles,
has a long story. Most popular approach in this direction is
the formulation of the Lagrangian on the space-time,
extended by the anticommuting variables,
which upon quantization provide the system with the nontrivial spin.
This approach is closely related with the supersymmetric field theories
and is essential for the formulation of superparticle systems.

There is another, pure bosonic, approach in which  the Lagrangian
is formulated on the direct product of the
initial space-time by some orbit of Poincar\'e group.
The actions of this sort are, in fact, the particle counterparts
of  Born-Infeld type systems. This approach seems to be interesting
due its visible relation with orbit method of Kirillov-Konstant-Souriau
\cite{souriau}. Most developed investigation of such systems has been
presented in Refs. \cite{tomsk}.
Notice that the  bosonic approach is the only correct
in  $(2+1)$-dimensional systems, due to anyonic nature of planar particle.

The bosonic approach  admits an aesthetically attractive modification,
where  the  additional bosonic variables are encoded
in the dependence of the Lagrangian on higher-order derivatives.
These  systems seem to be interesting not only for their clear geometrical
meaning. The  higher-derivative parts in their
Lagrangians can be generated by some
field-theoretical mechanism, e.g. arising as quantum corrections.
Investigation of such sort of particle systems
became popular after remarkable work of
Polyakov \cite{polac}, where he  show, that evaluation of
the effective action of $CP^1$ model minimally coupled to the
 Chern-Simons field for the charged solitonic excitation
results in the action
\begin{equation}
{\cal S}_{eff} =\int (m + {\pi\over 2\theta}K_2)|d{\bf x}|,
\label{polac}\end{equation}
where $K_2$ is  world-line's torsion and $\theta$ is field coupling
 strength.

Later it was found, that this system describes  anyonic  analog of
Majorana field equations \cite{3d}.
Due to further  studies it became a part of physical folklore
that  relativistic systems can get nontrivial spin,
 if one adds to the Lagrangians the  higher-derivatives terms
(more precisely, the terms, depending on
 reparametrization invariants (extrinsic  curvatures) of world-line).
Some significant observations were done in connection with this
 subject, particularly, in the  description of three- and four- dimensional
particle with Majorana spectrum \cite{maj},
 $4d$ massless particles \cite{pl}.
The relation of the mentioned massless particle  model
 with $W-$ algebras has also been established \cite{rr}.

However, such  systems were not  studied completely
 even in the four- dimensional space-time,
where only the first and second extrinsic curvatures  were
considered. \\
In this note we attempt to fill this gap and
consider the simplest four-dimensional analog of (\ref{polac})
given by the action \cite{pc}
\begin{equation}
{\cal S} =\int(c_0+cK_3)d{\tilde s},\quad
d{\tilde s}=|d{\bf x}|\equiv sd\tau\neq 0
\label{0}\end{equation}
where $\tau$ is an arbitrary evolution parameter
and $K_3$ is  the  torsion (highest curvature) of a worldline in $4d$ space.

We will show that this system  possesses interesting properties
which make it drastically different from other
 four- and three- dimensional massive particle systems depending
on extrinsic curvatures:
\begin{itemize}
\item it  has a zero spin;
\item it possesses, in addition to reparametrization invariance, the
 extra gauge degrees of freedom: its classical trajectories are
restricted by the condition
$$
 \frac{K^2_2K_3}{K^2_1}=|\alpha|, \quad \alpha={c_0}/{c}
$$
while the mass spectrum is described
the conformal
mechanics with the energy ${\cal E}$,
$$ dq\wedge dp, \quad {\cal H}=\frac{p^2}{2}\pm\frac{\alpha^2}{2q^2},
\quad
-2{\cal E}/\alpha^2= \left\{
\begin{array}{cc}
M^2/c^2_0\mp 1,&
 {\rm if}\;\;\alpha < 0,\\
\pm M^2/c^2_0 \pm 1,&
 {\rm if}\;\;\alpha >0
\end{array}\right. ,
$$
where
$q=K_1/K_2$, and  $M$ denotes the mass of the system.\\
When  $\alpha<0$, the upper sign corresponds to the
time-like trajectories, while other solutions corresponds
to the space-like ones.

\item  When $\alpha<-1/2$, the solutions
with space-like trajectories  possess  a {\it discrete} spectrum with
 massive and tachionic sectors,
while the solutions with time-like trajectories possess continuous
spectrum  containing massive, massless, and tachionic solutions.
%
\end{itemize}
The paper is arranged as follows:

In {\it Section 2} we construct the Hamiltonian system corresponding
to the model under consideration (\ref{0}).
We show that the system possesses an extended gauge invariance and
a zero spin.
The  geometry  of its classical trajectories and quantum spectrum
(in the Euclidean space) are  defined by the one-dimensional
conformal mechanics (with a repulsive potential).

In {\it Section 3}  we reformulate
 the Hamiltonian system constructed in Section 2 in the Minkowski space
     and consider the properties of its mass spectrum.

\setcounter{equation}{0}
\section{Hamiltonian  formulation}
In this section we give
the Hamiltonian formulation of the model (\ref{0}).

 Recall
that the extrinsic curvatures $K_a$ of a (non-null) curve in
four-dimensional
space can be defined via the Frenet equations for the moving frame
${\bf e}_a$ :
\begin{eqnarray}
&{\dot{\bf x}}=s{\bf e}_1,\quad {\dot{\bf e}}_a=k_a^{\;b}{\bf e}_b,
\quad{\bf e}_a{\bf e}_b=\eta_{ab},\quad a,b,c=1,2,3,4; &\label{f}\\
& k_{a}^{\; c}\eta_{cb}=
\left(
\begin{array}{cccc}
0   &k_1 &0   &0\\
-k_1&0   &k_2 &0\\
0   &-k_2&0   &k_3\\
0   &0   &-k_3&0
\end{array}\right),\quad k_{a-1}=sK_{a-1},& \label{mat}
\end{eqnarray}
 where   ${\bf x}$  are  coordinates of four-dimensional space,
and ${\bf e}_a$ are  elements of the moving frame.\\
While $K_1, K_2$ are positive quantities, the sign of the
highest curvature $K_{3}$ (torsion) is not  uniquely defined
by the Frenet equations \cite{postnicov}.
Without loss of generality we assume  below that  $K_3>0$.

 In the  Euclidean space $\eta_{ab}=\delta_{ab}$, so the
 Frenet equations read
\begin{equation}
 {\dot{\bf e}}_a=k_a{\bf e}_{a+1}-k_{a-1}{\bf e}_{a-1},\quad
{\bf e}_{0}={\bf e}_5\equiv 0.
\label{ffe}\end{equation}

One can transform the Frenet equations in the Euclidean space
 into those in the Minkowski space,
performing
 the  following transition
\begin{equation}
({\bf e}_{\underline a},\; k_{\underline a},\;
k_{{\underline a}-1}, s)\to
(i{\bf e}_{\underline a},\; ik_{\underline a},\;
ik_{{\underline a}-1},(-i)^{\delta_{1{\underline a}}}s)
\label{tr}\end{equation}
 for some index  ${\underline a}$.   \\
Indeed, this transformation preserves the form of the
matrix in (\ref{mat}),
while  the element ${\bf e}_{\underline a}$ becomes time-like:
${\bf e}^2_{\underline a}=-1$.
So, we can give our basic derivations for the Euclidean case,
 reformulating them for the Minkowski space
  for a final analysis.

  It follows  from  (\ref{ffe})
that
\begin{equation}
s=\sqrt{{\dot{\bf x}}^2},\quad
k_a={\dot{\bf e}}_a{\bf e}_{a+1}=\sqrt{{\dot{\bf e}_a}^2-k^2_{a-1}}.
\label{cf}\end{equation}
Thus, the  Lagrangian appearing in the action (\ref{0})
in the {\it Euclidean} space
can be  replaced   by  the following
one
\begin{equation}
 L=c_0s+c\sqrt{{\dot{\bf e}}_3^2-k^2_2}+{\bf p}({\dot{\bf x}}-s{\bf e}_1)+
\sum_{i}{\bf p}_{i-1}({\dot{\bf e}}_{i-1}-k_{i-1}{\bf e}_{i}+
k_{i-2}{\bf e}_{i-2})-\sum_{i,j}d_{ij}({\bf e}_i{\bf e}_j-\delta_{ij}),
\label{l}\end{equation}
where
${\bf x}, {\bf p},{\bf e}_i,{\bf p}_{i-1}, s, k_i, d_{ij}$
are independent variables,
$i,j=1,2,3$.\\

Now we can perform the Legendre transformation for
this Lagrangian
 (referring for
details to \cite{tmp}).\\
The variables  ${\bf p}_{i-1}$ represent the momenta
conjugated to
 ${\bf e}_{i-1}$,
whereas the momenta conjugated to $(s, k_{i-1}, d_{ij})$
lead to the trivial constraints
\begin{equation}
 p^s\approx 0,\quad p^{i-1}\approx 0,\quad p^{ij}\approx 0.
\label{erunda}\end{equation}
Setting  $k_3\neq 0$  we find that the momentum
conjugated to ${\bf e}_3$ is of the form
\begin{equation}
{\bf p}_3=c\left({\dot{\bf e}}^2_3-k^2_{2}\right)^{-1/2}
{\bf{\dot e}}_3.
\label{pN}\end{equation}
Taking into account  (\ref{cf}), we get the  constraints
\begin{equation}
{\bf p}_3{\bf e}_{3}\approx 0,\quad
{\bf p}_3{\bf e}_{1}\approx 0,
\quad{\bf p}^2_3- ({\bf p}_3{\bf e}_{2})^2-c^2\approx 0.
\label{PhiNN0}\end{equation}
Then the construction of primary
Hamiltonian system becomes straightforward.

To simplify  the resulting system, one can stabilize
   trivial primary constraints
(\ref{erunda}) and exclude
them  from our considerations, which makes
the variables $s, k_{i-1}, d_{ij}$  lagrangian multipliers.
 Without
loss of generality  one can
also impose the gauge conditions (see for a details \cite{tmp})
$${\bf p}_3{\bf e}_2\approx 0,\quad
 {\bf p}_{2}{\bf e}_{2}\approx 0, \quad
{\bf p}_{2}{\bf e}_{1}\approx 0. $$

After these manipulations we get
the   Hamiltonian system
\begin{equation}
\begin{array}{c}
\omega=d{\bf p}\wedge d{\bf x}+
\sum_{i} d{\bf p}_i\wedge d{\bf e}_i, \\
\;\;\\
{\cal H}=s({\bf p}{\bf e}_1 -c_0)+\sum_{i} k_{i-1}\phi_{i-1.i}+
{k_3\over 2c}({\bf p}^2_3-c^2)+\sum_{i,j}d_{ij}
({\bf e}_i{\bf e}_j-\delta_{ij}),
\end{array}
\label{ss}
\end{equation}
with  primary constraints
\begin{eqnarray}
&{\bf p}_i{\bf e}_{j}\approx 0,\quad
i\geq j &\label{primary}\\
&{\bf e}_i{\bf e}_j-\delta_{ij}\approx 0, &
\label{u}\\
&{\bf p}{\bf e}_1 -c_0\approx 0,&\label{phi0}\\
&\phi_{i-1.i}\equiv{\bf p}_{i-1}{\bf e}_{i}-
{\bf p}_{i}{\bf e}_{i-1} \approx 0,
\quad{\bf p}^2_3-c^2\approx 0,&\label{phi}
\end{eqnarray}
where variables $s,k_i, d_{ij}$ play the role of
Lagrangian multipliers.\\

Let us stabilize the primary constraints (\ref{primary})-(\ref{phi}).\\
Stabilization of  (\ref{primary}) gives the following fixation
of the Lagrangian multipliers $d_{ij}$:
\begin{equation}
2d_{i.j}=k_3c\delta_{i.3}\delta_{j.3}-sc_0\delta_{1.i}\delta_{1.j},
\label{d}\end{equation}
so that  the equations of motion read
\begin{equation}
\begin{array}{c}
 \dot{\bf x}={\bf e}_1,\\
{\dot{\bf e}}_1=k_1{\bf e}_2,\\
 {\dot{\bf e}}_2=-k_1{\bf e}_1+k_2{\bf e}_3,\\
 {\dot{\bf e}}_3=-k_2{\bf e}_2+k_3{\bf p}_3/c\\
 {\dot{\bf p}}_3=-k_3c{\bf e}_3-k_2{\bf p}_2,\\
{\dot{\bf p}}_2=-k_1{\bf p}_1+k_2{\bf p}_3,\\
{\dot{\bf p}}_1=-s{\bf p}+k_1{\bf p}_2+sc_0{\bf e}_1, \\
{\dot{\bf p}}=0.
\end{array}\label{hem}\end{equation}
Stabilizing the remaining primary constraints,
 we get the following first-stage secondary constraints
\begin{equation}
{\bf p}{\bf e}_2\approx 0,
\quad{\bf p}_{1}{\bf e}_{3}\approx 0,
\quad {\bf p}_3{\bf p}_2\approx 0.
\label{secondary}\end{equation}
From the  ortogonality of (${\bf e}_i, {\bf p}_3/c$) to ${\bf p}_2$,
which follows  from (\ref{phi}),(\ref{primary}),(\ref{secondary})
 we conclude that
\begin{equation}
{\bf p}_2\approx 0.
\label{secondary1}\end{equation}
Stabilizing the first and second  constraints from (\ref{secondary})
and  the constraint (\ref{secondary1}), we get
 \begin{eqnarray}
&k_2=qk_1, \quad k_3=sc_0/cq^2, &\label{multiplyers}\\
& {\bf p}_1=q{\bf p}_3,&\label{tertiary}
\end{eqnarray}
 where
\begin{equation}
1/q\equiv{{\bf p}{\bf e}_3\over c_0}\neq 0.
\end{equation}
Then we get
\begin{equation}
 {\bf p}=c_0{\bf e}_1+c_0{\bf e}_3/q+
p{\bf p}_3,\label{p}
\end{equation}
where
\begin{equation}
 p\equiv {\bf p}{\bf p}_3/c^2.\label{pi}
 \end{equation}
Consistency of the equations of motion (\ref{hem}) with the Frenet equations
implies that
\begin{equation}
{\bf e}_a=\{{\bf e}_i,\;{\bf e}_4\equiv{\bf p}_3/c\}:
\quad {\bf e}_a{\bf e}_b= \delta_{ab}.\label{on}\end{equation}
Therefore, the initial Hamiltonian system can be formulated purely in
terms of the moving frame ${\bf e}_a$, coordinates ${\bf x}$, and momentum
${\bf p}$, while the relations (\ref{on}), (\ref{p})
play the role of constraints.

Taking into account the  constraint (\ref{p})),
we get the
expressions for the rotation generators:
\begin{equation}
{\bf J}={\bf p}\vee{\bf x}+\sum_{i}{\bf p}_i\vee{\bf e}_i=
{\bf p}\vee ({\bf x}-q{\bf p}_3/c_0)\label{J}\end{equation}
and the Casimirs
\begin{equation}
{\bf p}^2=(cp)^2+c^2_0(1+1/q^2),\quad
{\bf W}^2\equiv ({\bf p}\vee {\bf J})^2=0.
\end{equation}
Thus, {\it the system possesses a  zero spin
despite the dependence of the initial Lagrangian
on higher derivatives.}\\

The expression for rotation generators (\ref{J}) hints us to introduce the
``effective" coordinate
\begin{equation}{\bf X}\equiv {\bf x}-q{\bf p}_3/c_0\;:
\quad {\bf{\dot X}}=s{\bf p}/c_0,\quad {\bf p}=const
\label{X}\end{equation}
whose evolution equations look similar to the convenient relation
 between velocity and momentum of a massive particle
(note, that in spite of these relations
the mass of the system is not equal to $c_0$).

The reduction of the initial  Hamiltonian system (\ref{ss})
  by the constraints
 (\ref{on}), (\ref{p}), (\ref{tertiary}) leads to
 the following unconstrained system
\begin{equation}
 \omega^{red}= d{\bf p}\wedge d{\bf X}+\frac{c^2}{c_0}
{dp\wedge dq},\quad
{\cal H}^{red}=\frac{sc^2}{c_0}\left(
 p^2/2+\frac{c^2_0}{2c^2q^2}-\frac{{\bf p}^2- c^2_0}{2c^2}\right).
\label{fd}\end{equation}
The external curvatures of the system are related with the modular
``coordinate"  $q$ as follows
\begin{equation}
 \frac{K_2}{K_1}=q, \quad K_3=\frac{c_0}{cq^2},\quad\Rightarrow\quad
\frac{K^2_2K_3}{K^2_1}=\frac{c_0}{c}.
\label{Kq}\end{equation}
So, {\it  the system under consideration
possesses extended gauge invariance
 since the trajectories
 with the same ratio $K_{1}/K_{2}$ are gauge equivalent.}\\

Reducing the system (\ref{fd}) by ${\bf p}$, to exclude
the  trivial part of  dynamics (\ref{X}),
we get that the
evolution of the parameter $q$ is described by the textbook
mechanical system \cite{ll}
\begin{equation}
dp\wedge dq,\quad
{\cal H}_0=p^2/2+\frac{c^2}{2c^2_0q^2},\quad {\cal E}=
\frac{{\bf p}^2-c^2_0}{2c^2}.
\label{cm}\end{equation}
This  system  can be immediately integrated  at the classical
level
$$ q^2=\frac{({\bf p}^2-c^2_0){\tilde s}^2}{2c^2}+
\frac{2}{{\bf p}^2-c^2_0},\quad{\bf p}^2>c^2_0,$$
as well as at the quantum one.

Thus,  adding of the torsion term increases the absolute value of momenta
 of the system with respect to the torsionless system,
${\bf p}^2\geq c^2_0$.
The system (\ref{cm}) is known in literature as a conformal mechanics,
 due to the symmetry of its  action under conformal  transformations
generated  by
\begin{equation}
{\cal H}_0=p^2/2+\frac{c^2}{2c^2_0q^2},
\quad{\cal D}=p q,\quad  {\cal K}={q^2}/{2}\;.
\end{equation}
The ``energy"  spectrum is  continuous and has the lowest bound ${\cal E}=0$
which is not normalizable.
In \cite{nc} the conformal symmetry of this system
has been used for solving the problem of the ground state.

Recently this mechanism has been found to be adequate
to the problem of motion of charged  particle in the field
of  a charged black hole \cite{k}. Due to this
 observation the interest to the conformal mechanics
and its supergeneralizations \cite{pashnev},
 is renewed in the context of
study of probe  $D0$-brane dynamics in the external $D$-brane  field
(see e.g.\cite{branes} and refs therein).
Conformal mechanics arises in our model  a different context:
it defines  local gauge symmetry of the  particle system.

\setcounter{equation}{0}
\section{Transition to Minkowski space}

In the previous Section  we constructed the constrained Hamiltonian system
corresponding to the action (\ref{0}) in the Euclidean space.
We found that the
evolution and  the geometry of this system are described in terms of  the
non-relativistic  conformal mechanics with repulsive potential.
The purpose of this Section is to consider the relativistic aspects of
this system, i.e. to investigate the model (\ref{0}) in the
Minkowski space.
As we have mentioned in the beginning of Section 2,
 we can
use the results concerning the Euclidean case,
since the  Frenet formulae in the Euclidean space  can be transformed
 into the ones in
Minkowski space with the use of transition (\ref{tr}),
 where ${\underline a}$ denotes
an element of moving frame which becomes time-like, i.e.
${\bf e}^2_{\underline a}=-1$.
Correspondingly,
for the transition to {\it time-like} trajectories we have to choose
 ${\underline a}=1$, while for the transition to
 {\it space-like} trajectories
  we have to choose ${\underline a}\neq 1$.
 We do not consider here systems with {\it light-like} trajectories.

For convenience we will use the notation
\begin{equation}
{\bf p}^2\equiv-M^2,\quad \alpha=\frac{c_0}{c},
\end{equation}
so that $M^2>,=,< 0$  corresponds to
the massive, massless, and tachionic sectors of the system, respectively.

To reformulate the system under consideration  in
 the Minkowski space we have to supply  the transition
 (\ref{tr})  with appropriate
 transformations of characteristic constants $c_0, c$ and momenta
${\bf p}_i$ which  preserve the form of
the initial action (\ref{0})
 and the initial Hamiltonian system, namely,
\begin{equation}
\begin{array}{cccc}
{\underline a}=1:&({\bf e}_1,{\bf p}_1, k_1, s, c_0)&\to &
( i{\bf e}_1, -i{\bf p}_1, ik_1,-is, ic_0);\\
{\underline a}=2:&({\bf e}_2,{\bf p}_2, k_2, k_1)&\to &
( i{\bf e}_2, -i{\bf p}_2, ik_2, ik_1),\\
{\underline a}=3:&
({\bf e}_3,{\bf p}_3, k_3, k_2, c)&\to &
( i{\bf e}_3, -i{\bf p}_3, ik_3, ik_2, -ic),\\
 {\underline a}=4:&
(c, k_3)&\to &( -ic,  ik_3).
\end{array}
\label{htr}\end{equation}
Indeed, it is easy to see that it is only the
ortonormality condition
${\bf e}_a{\bf e}_b=(-1)^{\delta_{{\underline a}a}}\delta_{ab}$
that is changes in (\ref{ss}) under this transformation.
Consequently, the reduced system (\ref{cm}) with  appropriately
changed parameters describes the effective Hamiltonian system corresponding
to the action (\ref{0}) in the Minkowski space.

The transition  (\ref{htr}) induces the following
transformation of the parameters and coordinates
of the reduced system (\ref{cm})
\begin{equation}
(p,q,\alpha, c_0, s)\to
\left\{
\begin{array}{cc}
(p, iq, i\alpha, ic_0, -is)&{\rm if } \;{\underline a}=1\\
(p,q,\alpha, c_0, s)& {\rm if } \;{\underline a}=2\\
(ip, -iq, i\alpha, c_0, s ) &{\rm if } \;{\underline a}=3\\
(-p, q, i\alpha, c_0, s ) &{\rm if } \;{\underline a}=4.
\end{array}\right.
\end{equation}
Thus,  the reduced system  reads
\begin{equation}
dp\wedge dq,\quad
\epsilon_{{\underline a}}s\left(
\frac{p^2}{2} +
\frac{\epsilon_{{\underline a}}\alpha^2}{2q^2}
-{\cal E}_{\underline a}\right)\approx 0,
\end{equation}
 where
$$
{\cal E}_{{\underline a}}=
-(-1)^{\delta_{4{\underline a}}}\frac{\alpha^2}{2}
({M^2}/{c^2_0}+(-1)^{\delta_{1{\underline a}}}),\quad
\epsilon_{{\underline a}}=\left\{
\begin{array}{cc}
1,& {\rm if }\;\;{\underline a}=1,2\\
-1,& {\rm if }\;\;{\underline a}=3,4.
\end{array}\right.
$$
The
expressions for curvatures (\ref{Kq}) read
\begin{equation}
 \frac{K_2}{K_1}=(-1)^{\underline a}q, \quad
K_3=(-1)^{\underline a}\frac{\alpha}{q^2},\quad\Rightarrow\quad
\frac{K^2_2K_3}{K^2_1}=(-1)^{\underline a}\alpha .
\label{Kqm}\end{equation}

Due to  positivity of curvatures $K_i$, we conclude that
\begin{itemize}
 \item  ${\underline a}=1,3$, $\quad q<0$,\hspace{0.5cm} if\hspace{0.5cm}
 $\alpha<0$,
\item
${\underline a}=2,4$, $\quad 0<q$,\hspace{0.5cm} if\hspace{0.5cm} $\alpha>0$.
 \end{itemize}
Thus,  in both cases the system possesses the solutions described
by conformal    mechanics with a repulsive (${\underline a}=1,2$) and an
attractive (${\underline a}=3,4$) potentials.
The classical solutions of these  systems are of the
form
$$
q^2=\left\{
\begin{array}{cc}
{\cal E}_{\underline a}{\tilde s}^2+(-1)^{[{\underline a}]}
{\alpha^2}/{{\cal E}_{\underline a}},&{\rm if} \;\; {\cal E}\neq 0,\\
2\alpha {\tilde s},&{\rm if}\;\; {\cal E}=0,\;\;
{\underline a}=3,4
\end{array}\right.
$$
When the potential
is attractive (${\underline a}=3,4$),
the parameter ${\tilde s}$ is defined on the domain
$$
{\tilde s}\in\left\{
\begin{array}{cc}
] -|\alpha/{\cal E}|,\;\;|\alpha/{\cal E}|[, & {\rm if}\; {\cal E}<0,\\
]|\alpha/{\cal E}| ,\;\;\infty [,& {\rm if} \;{\cal E}>0.
\end{array}\right.
$$
While  the  quantum  spectrum of the mechanics  with a repulsive potential
is continuous, the spectrum of  conformal mechanics with an attractive
potential
can be both  continuous and discrete.
 The discrete spectrum corresponds to the
strongly attractive potential ($|\alpha|>1/2$)
and has an infinite number of energy levels \cite{pr}.\\

We first consider the systems with repulsive potential
\begin{equation}
dp\wedge dq,\quad
{\cal H}_0=\frac{p^2}{2} +
\frac{\alpha^2}{2q^2},\quad \frac{2{\cal E}}{\alpha^2}=\left\{
\begin{array}{cc}
(1-{M^2}/{c^2_0}),& {\rm if}\;\; \alpha<0,\;\;{\underline a}=0\\
-(1+{M^2}/{c^2_0}),&{\rm if}\;\; \alpha>0,\;\;{\underline a}=1
\end{array}\right.
\end{equation}
Notice that while in the first case trajectories are {\it time-like},
those are {\it space-like}  in the second case. \\
Since the energy ${\cal E}$ is positive both in classical and quantum cases,
we get the restrictions on
the admissible value of the mass
\begin{itemize}
\item $M^2<c^2_0$,$\;$ for $\alpha<0$ and  time-like trajectories
(${\underline a}=1$),
\item  $M^2<-c^2_0$,$\;$ for $\alpha>0$, ${\underline a}=2$.
\end{itemize}
Thus,  in the first case there are massive, massless, and tachionic
states, while in the second case the system is tachionic.\\

Now we consider the systems with an attractive potential. In this case
all trajectories are space-like and are defined by the mechanics
\begin{equation}
dp\wedge dq,\quad
{\cal H}_0=\frac{p^2}{2} -
\frac{\alpha^2}{2q^2},\quad \frac{2{\cal E}}{\alpha^2}=
\left\{
\begin{array}{cc}
-(1+{M^2}/{c^2_0})&
{\rm if}\;\; \alpha<0,\; {\underline a}=3\\
(1+{M^2}/{c^2_0})&
{\rm if} \;\;\alpha>0,\; {\underline a}=4
\end{array}\right.
\end{equation}
The spectrum of conformal mechanics with an attractive potential is
continuous    for  ${\cal E}>0$ and for  ${\cal E}\leq 0, \alpha^2<1/4$
(see \cite{ll}).

If the potential is ``strongly attractive" $\alpha^2>1/4$,
and  ${\cal E}<0$,
the system has a dicrete spectrum with an
infinite number of bound states  given
by the expression \cite{pr}
$$
{\cal E}_n=-\hbar^2 B^2\exp{\frac{-2\pi n}{\sqrt{\alpha^2-1/4}}},\quad
n=0,\pm1,\pm2,\ldots,
$$
where $B$ is an undefined "phase" factor.   \\
So,
\begin{equation}
M^2_n/c^2_0=-1+
(-1)^{{\rm sgn}\alpha}\left(\frac{B\hbar}{\alpha}\right)^2
\exp{\frac{-2\pi n}{\sqrt{\alpha^2-1/4}}},\quad |\alpha|>1/2.
\end{equation}
Therefore,  for $\alpha>0$ the  discrete branch corresponds to pure
tachionic states, while for $\alpha<0$ it contains massive, massless,
and tachionic  sectors. \\

%
%
{\large Acknowledgments.} I am  thankful to M.Plyushchay,
who  suggested  to investigate this  model and gave numerous advices
 and C.Sochichiu for valuable discussions and useful comments.

This  work has been partially supported by
grants INTAS-RFBR No.95-0829, INTAS-96-538 and INTAS-93-127-ext.

\end{document}